\documentclass[twocolumn]{aastex702}
\usepackage{seqsplit}

\submitjournal{ApJL}
\begin{document}

\title{A 32-day Quasi-periodic Modulation in the Post-peak Light Curve of the Superluminous Supernova SN~2018bsz}

\author[0009-0003-2609-3591]{Aiswarya Sankar.K}
\affiliation{Graduate Institute of Astronomy, National Central University, 300 Jhongda Road, 32001 Jhongli, Taiwan}
\email{d1129601@astro.ncu.edu.tw}

\author[0000-0002-1066-6098]{Ting-Wan~Chen} 
\affiliation{Graduate Institute of Astronomy, National Central University, 300 Jhongda Road, 32001 Jhongli, Taiwan}
\email{twchen@astro.ncu.edu.tw}

\author[0000-0003-1325-6235]{Se\'{a}n~J.~Brennan}
\affiliation{Max Planck Institute for Extraterrestrial Physics, Giessenbachstra\ss e 1, 85748 Garching b. München, Germany}
\email{sbrennan@mpe.mpg.de}

\author[0000-0003-2191-1674]{Morgan~Fraser}
\affiliation{School of Physics, University College Dublin, Belfield, Dublin 4,
Ireland}
\affiliation{Centre for Space Research, University College Dublin, Belfield,
Dublin 4, Ireland}
\email{morgan.fraser@ucd.ie}

\author[0000-0003-2611-7269]{Keiichi~Maeda}
\affiliation{Department of Astronomy, Kyoto University, Kitashirakawa-Oiwake-cho, Sakyo-ku, Kyoto 606-8502, Japan}
\email{keiichi.maeda@kusastro.kyoto-u.ac.jp}

\author[0000-0001-8385-3727]{Thomas~Moore}
\affiliation{Space Telescope Science Institute, 3700 San Martin Drive, Baltimore, MD 21218, USA}
\email{tmoore@stsci.edu}

\author[0000-0002-2898-6532]{Sheng~Yang}
\affiliation{Institute for Gravitational Wave Astronomy, Henan Academy of Sciences, Zhengzhou 450046, Henan, China}
\email{sheng.yang@hnas.ac.cn}

\author[0000-0002-9928-0369]{Amar~Aryan}
\affiliation{Graduate Institute of Astronomy, National Central University, 300 Jhongda Road, 32001 Jhongli, Taiwan}
\email{amar@astro.ncu.edu.tw}

\author[0000-0003-1637-9679]{Dietrich~Baade}
\affiliation{European Organisation for Astronomical Research in the Southern Hemisphere (ESO), Karl-Schwarzschild-Str. 2, 85748 Garching b. München, Germany}
\email{dbaade@eso.org}

\author[0000-0003-0853-6427]{Ping~Chen}
\affiliation{Institute for Advanced Study in Physics, Zhejiang University, Hangzhou 310058, China}
\affiliation{Institute for Astronomy, School of Physics, Zhejiang University, Hangzhou 310058, China}
\email{chenp1220@gmail.com}

\author[0000-0001-8771-7554]{Chow-Choong~Ngeow}
\affiliation{Graduate Institute of Astronomy, National Central University, 300 Jhongda Road, 32001 Jhongli, Taiwan}
\email{cngeow@astro.ncu.edu.tw}

\correspondingauthor{Aiswarya Sankar.K}
\email[show]{d1129601@astro.ncu.edu.tw}  

\correspondingauthor{Ting-Wan~Chen}
\email[show]{twchen@astro.ncu.edu.tw}

\begin{abstract}
Recurrent structure in supernova (SN) light curves may reveal physical clocks associated with central-engine dynamics, binary orbital motion, or interaction with structured circumstellar material (CSM). We present a multiband photometric reanalysis of the nearby hydrogen-poor superluminous SN (SLSN-I) 2018bsz ($z=0.0267$), using \textit{Swift}/UVOT and GROND observations spanning approximately 110 rest-frame days after maximum light. After independently modelling and subtracting the smooth decline in each band, Generalized Lomb--Scargle periodograms reveal recurrent residual variations at broadly consistent phases across ten bands from $u$ to $K_s$. The strongest individual-band detections, in the $g$, $r$, $i$, $z$, and $J$ bands, yield periods of 31.5--31.8 days, while the joint multiband analysis gives a rest-frame period of $P=31.61^{+0.03}_{-0.03}$ days. The modulation is traced over approximately three cycles and is therefore described as quasi-periodic. The signal is robust to the adopted detrending procedure, while comparison-star and lunar-cycle tests disfavour an observational systematic. The modulation is broadly phase-coherent from the optical to the near-infrared, but its amplitude increases towards shorter wavelengths; lower-significance bands also show possible wavelength-dependent shifts in the best-fitting period. The present observations do not uniquely determine the physical origin. Lense--Thirring precession, interaction with CSM structured before the explosion, and post-SN interaction with a surviving companion remain possible. SN~2018bsz provides one of the clearest examples of coherent, month-scale photometric modulation in an SLSN-I.
\end{abstract}


\section{Introduction} \label{sec:intro}

Hydrogen-poor superluminous supernovae (SLSNe-I) form a spectroscopically distinct class of luminous stellar explosions. They span a range of peak luminosities, extending to approximately $10^{44}\,\mathrm{erg\,s^{-1}}$, and are substantially more luminous than ordinary core-collapse SNe \citep{quimby2011Natur.474..487Q,galyam2019ARA&A..57..305G}. Their power source remains debated, with proposed mechanisms including magnetar spin-down, radioactive decay, fallback accretion, and interaction between the SN ejecta and circumstellar material (CSM) \citep[see][for a review]{moriya2018SSRv..214...59M}. 

An important clue to the underlying power source is the prevalence of structure in post-peak SLSN light curves. A systematic study found that 44--76 per cent of well-observed SLSNe-I cannot be adequately described by a smooth magnetar-powered decline alone \citep{Hosseinzadeh2022ApJ...933...14H}. Examples include SN~2015bn, LSQ14an, SNe~2017gci, 2017egm, and 2020qlb, which exhibit variations ranging from low-amplitude undulations to prominent bumps and rebrightenings \citep{2015bn2016ApJ...826...39N,2017MNRAS.468.4642I,2021MNRAS.502.2120F,2023ApJ...949...23Z,2020qlb2023A&A...670A...7W}. Proposed explanations include low-level variability of the central engine, small changes in ejecta opacity, and interaction with confined or structured CSM \citep{Hosseinzadeh2022ApJ...933...14H,2017ApJ...848....6Y,2018ApJ...867L..31C,2023ApJ...943...42C}. Photometry alone rarely distinguishes uniquely among these possibilities. However, recurrent variations are particularly informative because they may reveal an underlying physical clock associated with central-engine dynamics, binary orbital motion, or periodically structured CSM.

Recurrent light-curve modulations have been reported in a small number of stripped-envelope SNe. The Type~Ic SN~2022jli displayed 12.4-day undulations with an amplitude of approximately $0.1$ mag across multiple bands for nearly 200 days. These were attributed to a post-explosion binary in which a compact remnant accretes material from a surviving companion during successive close passages \citep{thomas2023ApJ...956L..31M,ping2024Natur.625..253C,2026A&A...707A.161C}. SN~2022esa exhibited an approximately 32-day modulation over multiple cycles, interpreted most likely as ejecta interaction with periodically structured, oxygen-rich CSM associated with a Wolf--Rayet binary, although post-SN binary interaction remains possible \citep{keiichi2026PASJ...78L...1M}. Among SLSNe-I, SN~2015bn and SN~2020qlb showed only two to three undulations in their magnetar-model residuals, with characteristic timescales of approximately 30--50 days and $32\pm6$ days, respectively \citep{2015bn2016ApJ...826...39N,2020qlb2023A&A...670A...7W}. The limited number of cycles prevented strict periodicity from being established. More recently, the chirped light-curve modulations of SN~2024afav were attributed to Lense--Thirring precession of a fallback disk around a newborn magnetar \citep{2026Natur.651..321F}, while its spectra independently showed evidence for ejecta--CSM interaction \citep{kumar2026ApJ...998L...3K}. These events illustrate both the diagnostic potential of recurrent variability and the difficulty of identifying its physical origin.

SN~2018bsz is one of the nearest and best-observed SLSNe-I \citep[$z=0.0267$;][]{anderson2018bsz2018A&A...620A..67A}. It displayed a long pre-maximum plateau and unusually strong C~II absorption, followed by the emergence of multicomponent H$\alpha$ emission approximately 30 days after maximum. Its subsequent spectroscopic and polarimetric evolution has been interpreted as interaction with highly aspherical, possibly disk-like, hydrogen-rich CSM \citep{mikka2022A&A...666A..30P}. SN~2018bsz also formed a substantial mass of carbon dust within its metal-rich ejecta at late times \citep{janet2021arXiv210907942C}. Although the dust formation is not itself direct evidence for CSM interaction, it demonstrates the unusual complexity of the late-time ejecta. In addition, the relatively flat late-time X-ray light curve is more readily explained by ejecta--CSM interaction than by direct magnetar emission \citep{2026ApJ..1005..212A}. The spectroscopic, polarimetric, and X-ray evidence for interaction, together with the extensive multi-band photometric coverage, makes SN~2018bsz a particularly valuable event in which to investigate whether recurrent light-curve structure traces interaction with structured CSM or another underlying dynamical clock.

In this \textit{Letter}, we report recurrent modulation in the declining light curve of SN~2018bsz with a characteristic timescale of approximately 32 days. The variations can be traced at broadly consistent phases across ten photometric bands from $u$ to $K_s$ over approximately 110 rest-frame days after maximum light, with the strongest statistical evidence coming from the best-sampled GROND bands with a daily cadence. Because the observations cover only approximately three cycles, we do not regard them as establishing strict periodicity. We examine the statistical significance and robustness of the modulation, test for potential observational systematics, and discuss possible physical interpretations.

We adopt a flat $\Lambda$CDM cosmology with
$H_0=73\,\mathrm{km,s^{-1}\,Mpc^{-1}}$, $\Omega_m=0.27$, and
$\Omega_\Lambda=0.73$.

\begin{figure*}[ht!]
    \centering
    \includegraphics[width=1.0\textwidth]{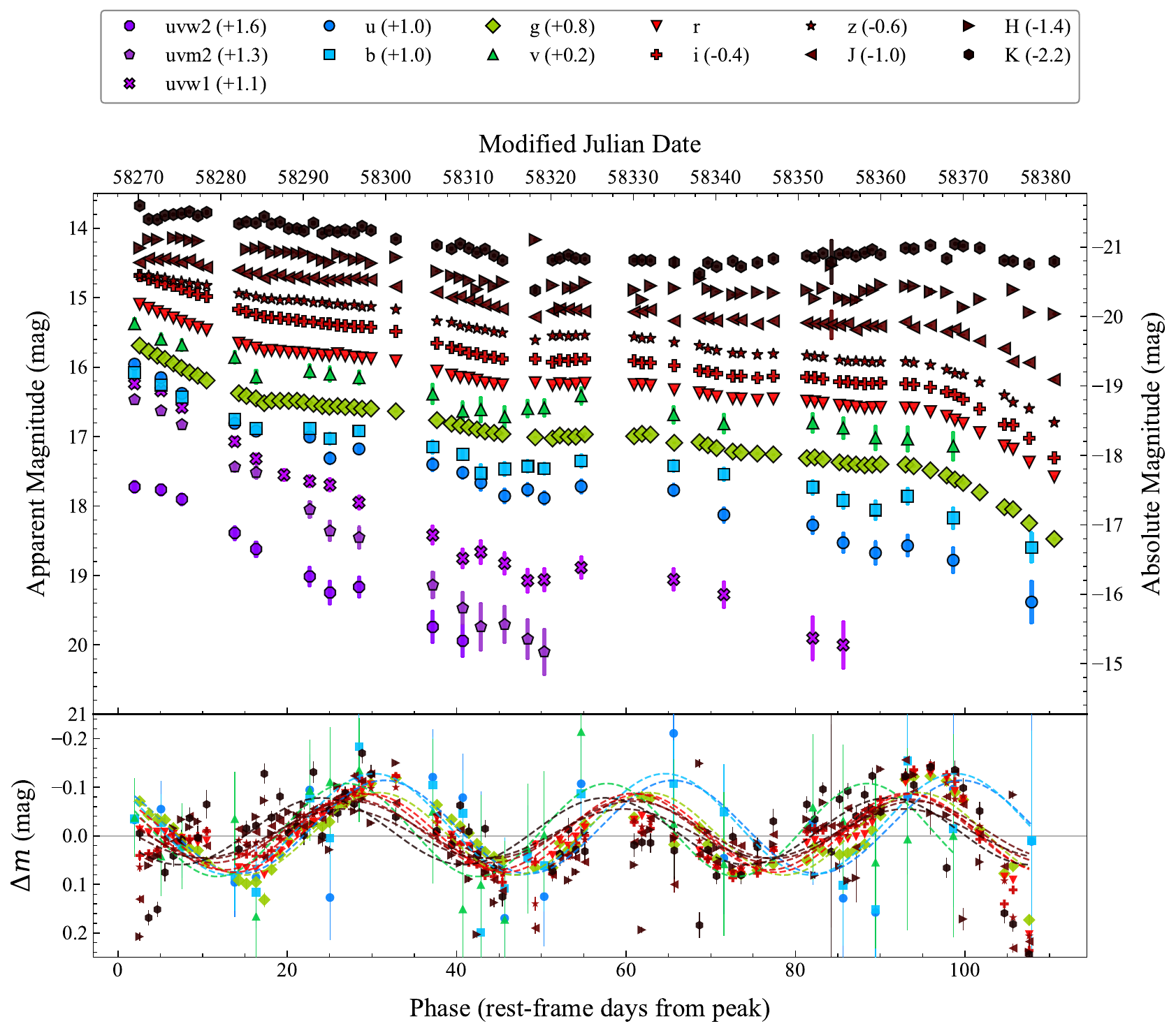}
    \caption{\textit{Top:} Multi-band light curves of SN~2018bsz. The left-hand $y$-axis corresponds to the observed apparent magnitudes, while the right-hand $y$-axis shows the absolute magnitudes after applying corrections for Milky Way extinction, host-galaxy extinction, and $K$-correction. The upper $x$-axis gives the Modified Julian Date (MJD), and the lower $x$-axis represents the rest-frame phase measured relative to the epoch of peak luminosity (MJD = 58267.5).
    \textit{Bottom:} Detrended light-curve residuals ($\Delta m$) as a function of rest-frame phase. The dashed curves show the best-fitting sinusoidal models based on the Lomb--Scargle periods for each photometric band.
}
    \label{fig:LC}
\end{figure*}

\begin{figure*}[ht!]
    \centering
    \includegraphics[width=0.95\textwidth]{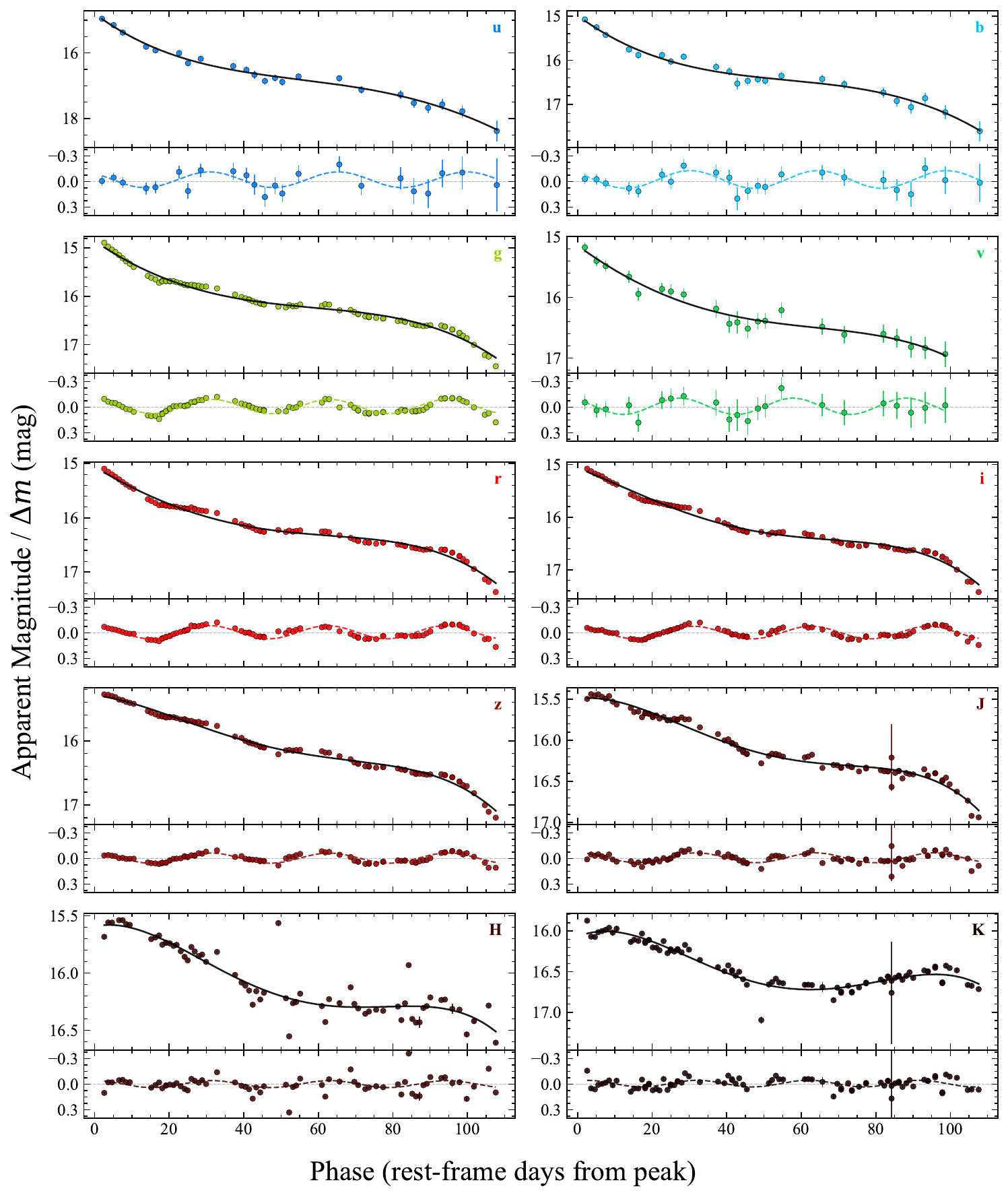}
    \caption{Detrending of the multi-band light curves of SN~2018bsz. \textit{Top:} Apparent-magnitude light curves in the ten photometric bands ($u$, $b$, $g$, $v$, $r$, $i$, $z$, $J$, $H$, $K_s$). The solid curves show the fourth-degree polynomial fitted independently to each band to trace the smooth underlying decline.
    \textit{Bottom:} Residuals $\Delta m$ obtained after subtracting the polynomial fit. The dashed curves show the best-fitting sinusoidal models based on the periods identified by the Lomb--Scargle analysis, illustrating the quasi-periodic modulation analyzed in Figure~\ref{fig:periodogram}.}
    \label{fig:detrending}
\end{figure*}

\section{Photometric Data} \label{sec:data}

In this work, we analyse the ultraviolet (UV), optical, and near-infrared (NIR) observations of SN~2018bsz originally presented by \citet{janet2021arXiv210907942C}. The UV and optical observations were obtained with the Ultraviolet and Optical Telescope (UVOT; \citealt{uvot2005SSRv..120...95R}) aboard the \textit{Neil Gehrels Swift Observatory} in the $uvw2$, $uvm2$, $uvw1$, $u$, $b$, and $v$ bands. Simultaneous optical and NIR observations were obtained with the Gamma-Ray Burst Optical/Near-Infrared Detector (GROND; \citealt{GROND2008PASP..120..405G}) mounted on the Max Planck Gesellschaft (MPG) 2.2-m telescope at European Southern Observatory's (ESO) La~Silla Observatory, covering the $g$, $r$, $i$, $z$, $J$, $H$, and $K_s$ bands. The combined data set spans rest-frame phases from $+2$ to approximately $+110$ days, where phase zero is defined as the $r$-band maximum at MJD~58267.5 \citep{anderson2018bsz2018A&A...620A..67A}.

For the present analysis, we applied updated photometric processing to these data. The GROND photometry was remeasured using an updated version of the Automated Photometry Of Transients pipeline (\textsc{AutoPhOT} \footnote{\url{https://github.com/Astro-Sean/autophot}}; \citealt{autophot2022A&A...667A..62B}), while the Swift/UVOT photometry was recalibrated using the updated long-term sensitivity corrections from the Swift/UVOT Calibration Database (CALDB). The contribution from the underlying host galaxy was removed from all photometric measurements. All magnitudes are reported in the AB system.

\section{Analysis} \label{sec:analysis}

\subsection{Underlying light-curve evolution and detrending}
\label{sec:detrending}

The multi-band light curves of SN~2018bsz are shown in Figure~\ref{fig:LC}. 
We restrict the detrending and period-search analysis to the eleven sufficiently sampled $uvw1$, $u$, $b$, $g$, $v$, $r$, $i$, $z$, $J$, $H$, and $K_s$ light curves. The \textit{Swift}/UVOT $uvw2$ and $uvm2$ light curves are also included, but the sampling is too sparse to constrain both the underlying decline and the shorter-timescale residual variations reliably.
The full photometric data set extends from maximum light to approximately $+130$ rest-frame days, owing to the late-time Swift/UVOT $u$- and $b$-band observations. However, the dense multi-band coverage provided by the GROND 7-colour observations ends at approximately $+110$ rest-frame days. We therefore restrict the subsequent detrending and periodicity analyses to phases between maximum light and $+110$ days. The multi-band light curves are dominated by an overall post-peak decline, with lower-amplitude undulations superimposed on the underlying evolution. These undulations are visible across the GROND optical and NIR bands, although their amplitudes and statistical significance vary with wavelength.

To isolate these shorter-timescale variations, we fitted the light curve in each photometric band independently with a fourth-degree polynomial. The smooth evolution differs among the bands, particularly between the optical and NIR, and was therefore not represented using a common polynomial model. Similar polynomial detrending has been used in previous searches for periodic or
quasi-periodic SN variability \citep{ping2024Natur.625..253C,thomas2023ApJ...956L..31M,
keiichi2026PASJ...78L...1M}.

We define the residual magnitude as
\begin{equation}
    \Delta m = m_{\rm poly}-m_{\rm obs},
\end{equation}
such that positive values correspond to an excess in brightness relative to the smooth decline. The resulting residuals are shown in Figure~\ref{fig:detrending}. 

To assess the sensitivity of our results to the adopted model for the underlying light-curve evolution, we repeated the complete detrending and period-search procedure using third- and fifth-degree polynomials, in addition to the fourth-degree polynomial adopted in the main analysis. Both alternative polynomial orders preserve the recurrent structure in the residuals and recover a characteristic timescale close to (rest-frame) 32 days (Figures~\ref{fig:detrend3} and \ref{fig:detrend5}). As a further test, we replaced the polynomial baseline with a Bazin function and repeated the analysis. The resulting residuals likewise exhibit a similar sequence of maxima and minima and a characteristic timescale consistent with that obtained in the main analysis (Figure~\ref{fig:detrend_bazin}). These tests indicate that the recovered modulation is not strongly dependent on either the adopted polynomial order or the particular functional form used to describe the underlying light-curve evolution.

The peaks and troughs in the residuals occur at broadly consistent phases across the optical and NIR bands. However, the modulation amplitude increases towards shorter wavelengths. The variations are therefore approximately phase-coherent across wavelength, but are not strictly achromatic.

\begin{figure}[ht!]
    \centering
    \includegraphics[width=0.95\columnwidth]{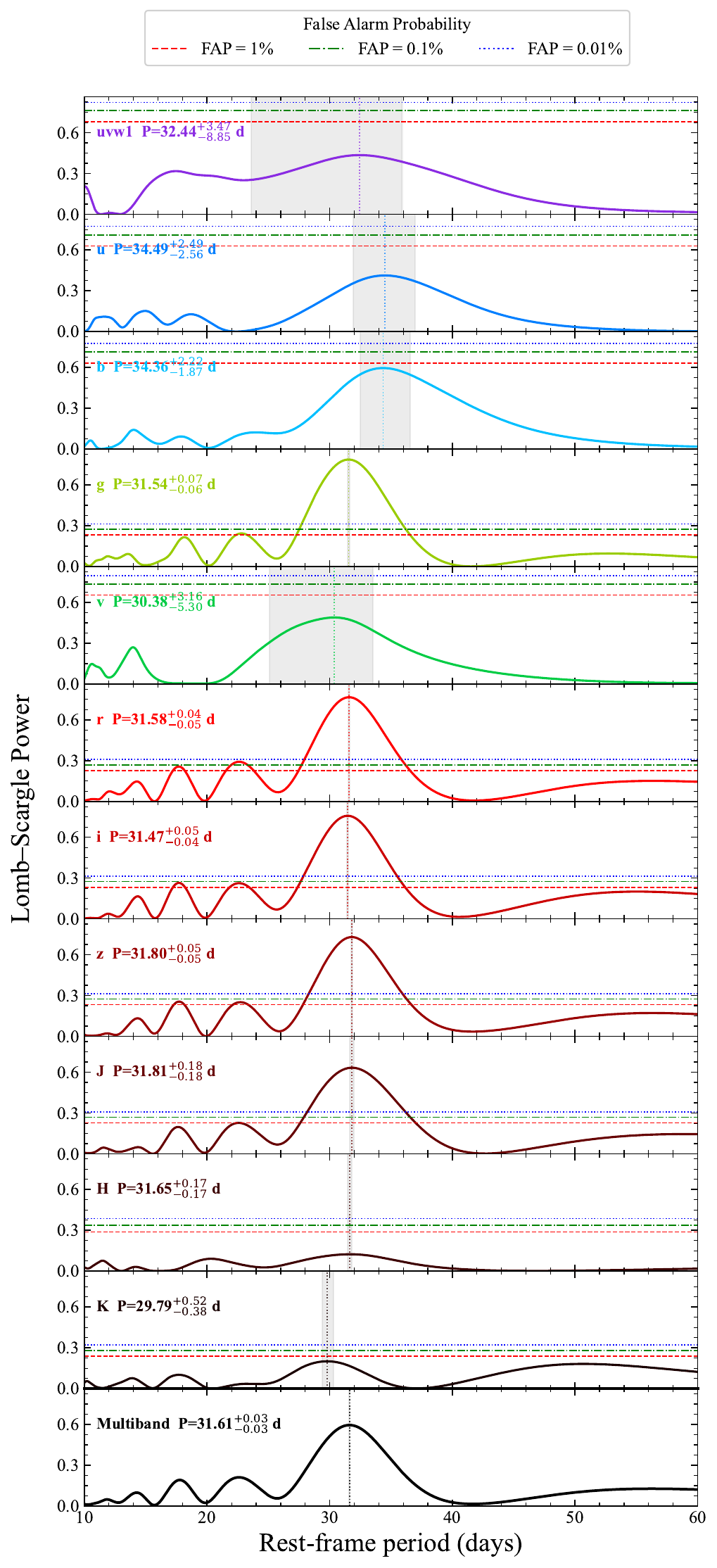}
    \caption{Single-band Lomb--Scargle periodograms are shown for all photometric filters, computed independently for each band to examine periodic signals present in the residual light curves with the period labeled for each band. In addition, a multiband Lomb--Scargle periodogram is computed using the \texttt{Astropy} implementation, combining all available photometric data from $uvw1$ through $K_s$. The multiband periodogram yields a dominant period of $P=31.6$ days. The dashed lines indicate the False Alarm Probability for each band. The grey shaded region shows the 16th--84th percentile range from the Monte Carlo procedure discussed in Section~\ref{sec:monte_carlo}.
    }
    \label{fig:periodogram}
\end{figure}

\subsection{Lomb--Scargle period search}
\label{sec:period_search}

We searched for recurrent signals in the detrended light curves using the Generalized Lomb--Scargle (GLS) periodogram \citep{GLS2009A&A...496..577Z,lombscargle2018ApJS..236...16V}, as implemented in \texttt{astropy.timeseries.LombScargle} \citep{astropy2022ApJ...935..167A}. Each photometric band was analysed independently over periods from $P_{\rm min}=10 $ to $P_{\rm max}=60$ days. The adopted limits were chosen based on the observational cadence and temporal baseline.

We also calculated a combined multiband periodogram using the \texttt{\seqsplit{astropy.timeseries.LombScargleMultiband}} with \texttt{method=`flexible'} \citep{multiband2015ApJ...812...18V,astropy2022ApJ...935..167A}. This method fits a common trial frequency to all bands while allowing band-dependent offsets and periodic components.

The individual and multiband periodograms are presented in Figure~\ref{fig:periodogram}. Several bands show their dominant peak at approximately 29--35 days. The peaks in the $g$, $r$, $i$, $z$, and $J$ bands exceed the 0.01 per cent false-alarm probability (FAP) level, whereas the remaining bands show lower-significance peaks at broadly consistent periods. The multiband periodogram yields a dominant period of $P=31.6$ days.

The analytic FAP levels of 1, 0.1, and 0.01 per cent shown in
Figure~\ref{fig:periodogram} were calculated following
\citet{fap2008MNRAS.385.1279B}. These values assume the adopted
noise model and do not include uncertainty associated with the
choice of secular light-curve model. We therefore assess the
period uncertainty and empirical significance separately in
Section~\ref{sec:monte_carlo}.

Several red optical and NIR bands also show a broad secondary feature near approximately 60 days. Because this timescale is close to twice the dominant period and is sampled by fewer than two complete cycles, it is poorly constrained. It may arise from the finite baseline, observational sampling, or the adopted detrending procedure. We therefore do not interpret it as an independent periodicity. The $uvw1$ light curve shows a broad periodogram maximum at P=32.4 d, consistent with the optical modulation, but the period is poorly constrained and not formally significant; we therefore regard it as an independent consistency check rather than a standalone detection.

\subsection{Period uncertainties and Monte Carlo analysis}
\label{sec:monte_carlo}

We estimated the uncertainty on the recovered period using
Monte Carlo simulations. In each of $N=1000$
realizations, the measured magnitude at every epoch was perturbed
by a random value drawn from a Gaussian distribution with a
standard deviation equal to its photometric uncertainty. The
secular light curve was then refitted and subtracted, and the GLS
analysis was repeated using the same frequency grid as for the
observed data. Repeating the complete procedure ensures that the
quoted uncertainty includes the coupling between the secular
light-curve fit and the period search.

The recovered period distribution has a median of $P_{50}=31.61$~days, with 16th and 84th percentiles of $-0.03$ and $+0.03$~days, respectively.

To estimate an empirical false-alarm probability, we additionally
generated null light curves at the actual observing epochs,
preserving the measured photometric uncertainties but without an
injected periodic signal. Each simulated light curve was passed
through the same detrending and period-search procedure. The
empirical FAP was calculated as the fraction of simulations whose
maximum periodogram power equalled or exceeded that measured in
the observations. With zero exceedances among 10{,}000 null realizations, the empirical FAP is formally unresolved by the simulations. Using the exact single-sided binomial upper limit for zero events \citep{Gehrels1986ApJ...303..336G}, we obtain FAP $< 3.0\times10^{-4}$ at 95\% confidence.

The observations trace the coherent modulation over approximately three cycles. We therefore describe the signal as quasi-periodic; throughout the remainder of this paper, \textit {``period''} refers to its best-fitting recurrence timescale.

\begin{figure}[ht!]
    \centering
    \includegraphics[width=0.5\textwidth]{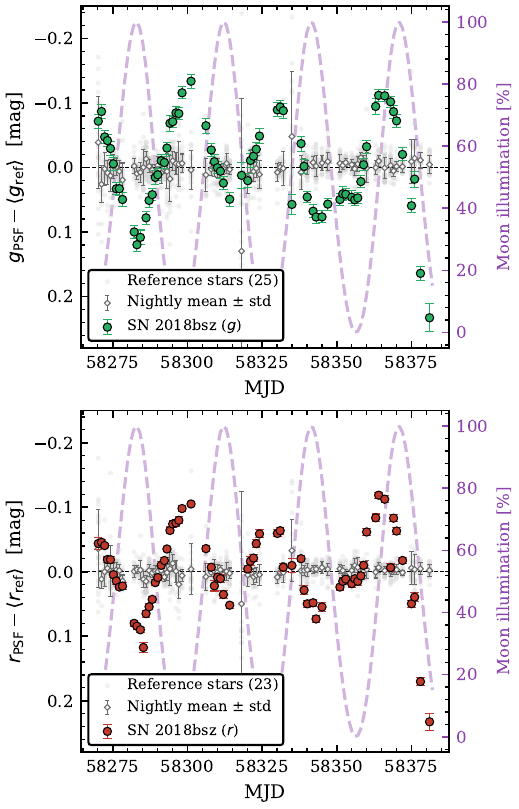}
\caption{
  Differential photometry of SN\,2018bsz in the $g$ and $r$ bands from GROND imaging (MJD\,$<$\,58400) in observed-frame.  Green and red circles show the residual of the SN\,2018bsz differential PSF magnitude, $g_{\rm PSF} - \langle g_{\rm ref}\rangle$, after subtraction of a 4th-order polynomial fit to the early-time light curve; error bars are the 1$\sigma$ PSF photometric uncertainties.  Grey points are the mean-subtracted magnitudes of 25 (23) non-variable reference stars in the $g$-( $r$)-band field, and open diamonds mark their nightly mean with error bars representing the nightly standard deviation. The purple dashed line shows the moon illumination in observed-frame with the percentage as shown in the right-hand $y$-axis. In the $g$-band residual of SN\,2018bsz ($\sim$0.072\,mag) exceeds the reference-star scatter ($\sim$0.036\,mag) by a factor of $\sim$2, indicating residual intrinsic variability on top of the smooth decline, and not an instrument effect or contamination from Lunar illumination.}
    \label{fig:moon_ref_stars}
\end{figure}

\subsection{Tests for observational systematics and moonlight contamination}
\label{sec:systematics}

A potential systematic concern for the ground-based photometry is contamination by scattered moonlight. The lunar synodic period is 29.53 days in the observer frame and  this is shorter than the recovered period of approximately 32 days. Over the approximately 110-day interval during which the modulation can be traced, the two signals would accumulate a relative phase shift of nearly half a cycle. 

For each GROND epoch, we calculated the fractional lunar illumination using \texttt{astropy.coordinates} \citep{astropy2022ApJ...935..167A}, adopting the built-in ephemeris and the coordinates of La~Silla Observatory. Fractional illumination alone is not a quantitative predictor of the lunar contribution to the sky background, which also depends on the lunar phase function, the Moon--target separation, airmass, atmospheric extinction, and wavelength \citep{Krisciunas1991PASP..103.1033K,Jones2013A&A...560A..91J}. We therefore use it only as a qualitative diagnostic. The maxima and minima of the SN residuals show no persistent correspondence with high lunar illumination.

Figure~\ref{fig:moon_ref_stars} compares the detrended $g$- and $r$-band residuals of SN~2018bsz with the mean-subtracted photometry of 25 and 23 comparison stars, respectively, measured in the same GROND images. For each band, the comparison stars were selected by requiring a detection with a signal-to-noise ratio of $S/N>5$ in every image; the resulting samples span a range of apparent magnitudes. The fractional lunar illumination at the GROND epochs is also overplotted. The SN residual maxima and minima show no persistent correspondence with the lunar illumination curve. After subtracting the mean magnitude of each star across all epochs, the nightly residuals of the 25 and 23 comparison stars in the $g$ and $r$ bands, respectively, remain close to zero and show no coherent pattern resembling the SN modulation. In the $g$ band, the standard deviation of the SN residuals is approximately $0.072$ mag, about twice that of the comparison-star residuals ($\sim0.036$ mag). These results argue against a common field-wide photometric zeropoint or sky-background systematic as the origin of the modulation.

Finally, variations at broadly consistent phases are present in the space-based \textit{Swift}/UVOT photometry. Because these observations are not subject to atmospheric scattering of moonlight, they provide an independent consistency check, although the significance in the individual UVOT bands is lower than in the best-sampled GROND bands. Taken together, the mismatch with the lunar period, the stability of the comparison stars, and the independent Swift/UVOT measurements make lunar contamination or a common photometric systematic an unlikely explanation for the observed modulation.

\section{Discussion}
\label{sec:discussion}

The analysis above reveals a coherent, quasi-periodic modulation with a period of approximately 32\,days in the optical and NIR light curves of SN~2018bsz. The signal can be traced over approximately three cycles, with its peaks and troughs occurring at broadly consistent phases across wavelength, although the modulation amplitude increases towards the blue. The recovered period and phasing remain consistent under alternative models of the underlying light-curve evolution, while the comparison-star and lunar-cycle tests disfavour observational systematics as the origin of the signal. In the following, we compare SN~2018bsz with other recurrently modulated SNe and consider possible physical origins of the modulation.

\subsection{Comparison with other recurrently modulated supernovae}
\label{sec:comparison}

\begin{figure}[ht!]
\centering
\includegraphics[width=0.5\textwidth]{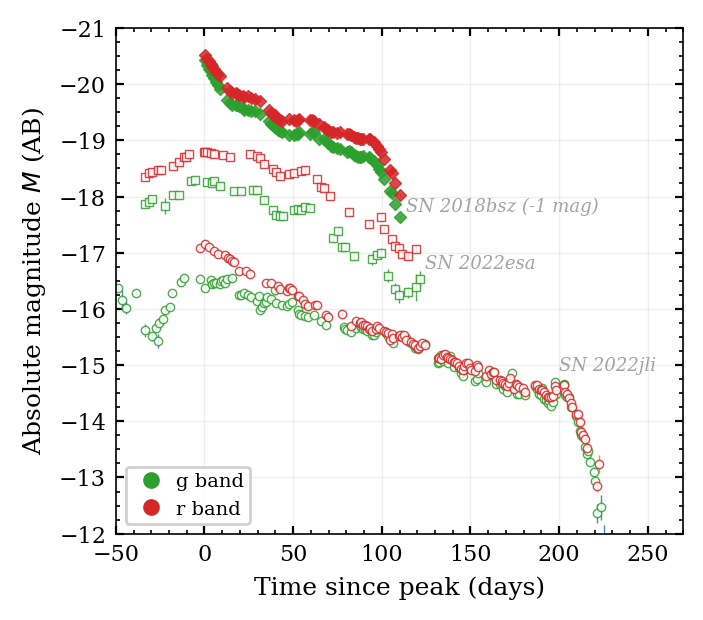}
\caption{\textit{Top:} Post-peak $r$ and $g$-band light curves of SNe~2022jli, 2022esa, and 2018bsz, aligned in rest-frame phase relative to peak, with offset as labeled.
}
\label{fig:lc_comparison}
\end{figure}

SNe with well-established recurrent optical light-curve modulations remain rare, and the best-observed examples show considerable diversity. SN~2022jli, a Type~Ic SN, exhibited 12.4-day modulations with a peak-to-peak amplitude of approximately 0.1\,mag over many cycles following its secondary light-curve maximum \citep{thomas2023ApJ...956L..31M,ping2024Natur.625..253C,2026A&A...707A.161C}. SN~2022esa, classified as an SN~Ic-CSM, showed an approximately 32-day modulation over about 200\,days, with a possible gradual increase in the period \citep{keiichi2026PASJ...78L...1M}. Its period is therefore close to that of SN~2018bsz, although it was observed over substantially more cycles. The post-peak $g$- and $r$-band light curves of SN~2018bsz, SN~2022jli, and SN~2022esa are compared in Figure~\ref{fig:lc_comparison}.

A recent systematic search of ZTF stripped-envelope SNe also identified SN~2020sgf as a promising candidate for an approximately 30-day recurrent modulation in both the $g$ and $r$ bands, although its periodic nature remains to be independently confirmed \citep{Horowicz2026arXiv260818207H}.

For SLSNe-I, SN~2024afav provides a contrasting case. Its post-peak modulations showed a pronounced decrease in both period and amplitude, with the period shortening by approximately $29\pm10$\% between successive cycles \citep{2026Natur.651..321F}. SN~2018bsz shows no evidence for such rapid chirping: its approximately three observed cycles are consistent with a constant characteristic timescale. However, the shorter temporal baseline of SN~2018bsz limits our sensitivity to modest period evolution. SLSNe-I such as SN~2015bn and SN~2020qlb also showed two or three approximately regularly spaced undulations, but their limited number of cycles likewise prevented confirmation of strict periodicity \citep{2015bn2016ApJ...826...39N,2020qlb2023A&A...670A...7W}.

\begin{figure}[ht!]
\centering
\includegraphics[width=0.5\textwidth]{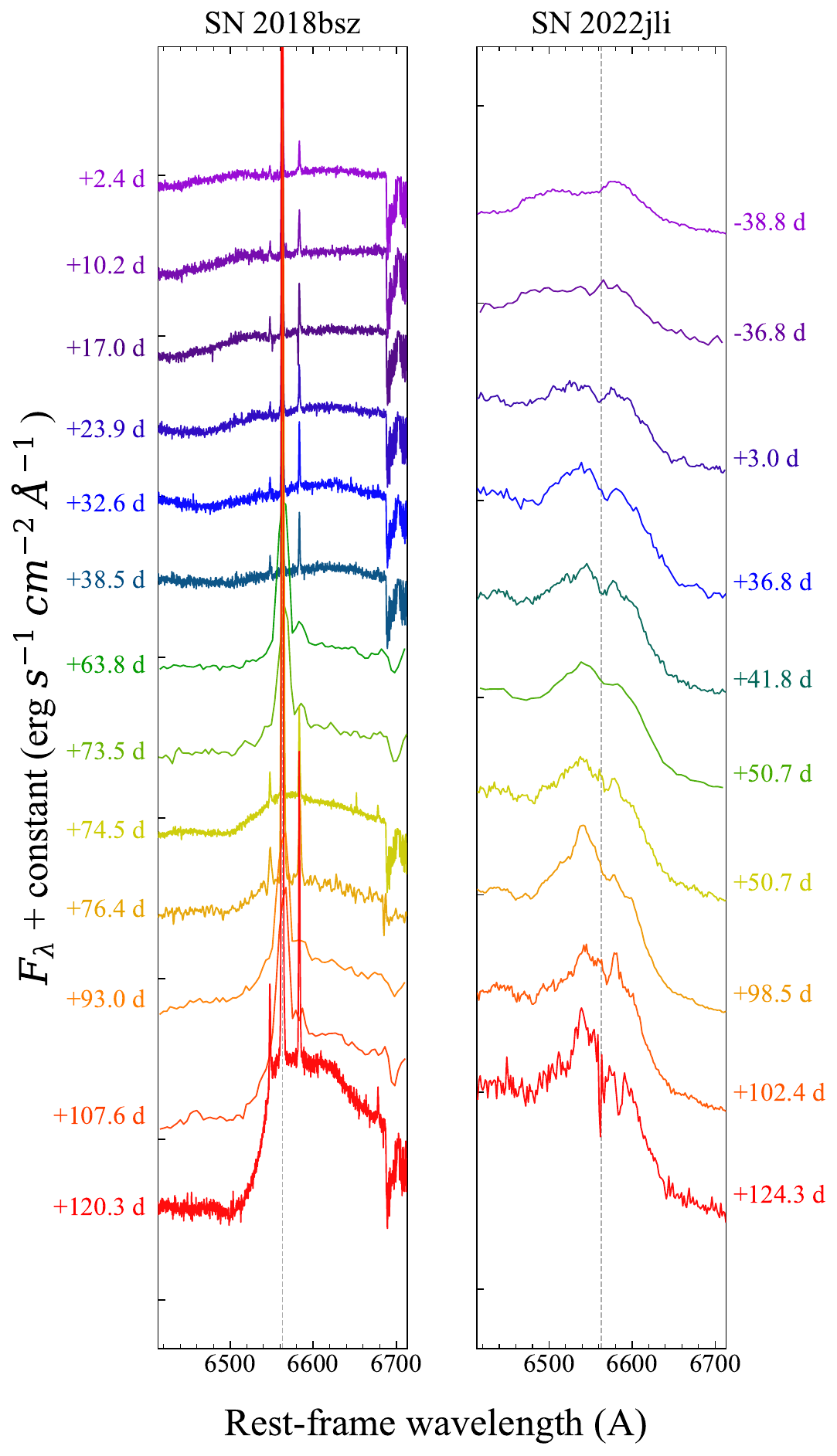}
\caption{Comparison of the H$\alpha$ line regions in SNe~2018bsz and 2022jli. Only the broad, SN-associated emission components are considered. The unresolved narrow H$\alpha$ emission in SN~2018bsz, which is present from the earliest spectra, is excluded because it is likely dominated by emission from the star-forming host galaxy rather than by the SN. SN~2022jli exhibits periodic radial-velocity shifts in its H$\alpha$ emission correlated with the light-curve modulation \citep{ping2024Natur.625..253C}, whereas no obvious shift in the centroid of the broad H$\alpha$ component is seen among the available spectroscopic epochs of SN~2018bsz.}
\label{fig:spectra}
\end{figure}

SN~2022jli also displayed periodic radial-velocity shifts in a narrow H$\alpha$ component, providing an additional diagnostic connected to its photometric modulation \citep{ping2024Natur.625..253C}. SN~2018bsz developed multi-component H$\alpha$ emission at approximately 30\,days after maximum \citep{mikka2022A&A...666A..30P}. Motivated by the behaviour of SN~2022jli, we examined the available spectra of SN~2018bsz \citep{mikka2022A&A...666A..30P,janet2021arXiv210907942C} for corresponding changes in its broad and intermediate-width H$\alpha$ and H$\beta$ components. We find no comparable, measurable displacement of their centroids among the available epochs (Figure~\ref{fig:spectra}). This is not a direct like-for-like comparison with the narrow H$\alpha$ component of SN~2022jli, and the sparse spectral sampling of SN~2018bsz does not exclude lower-amplitude or phase-dependent velocity variations.

\subsection{Physical interpretation}
\label{sec:physical_interpretation}

Several mechanisms could plausibly produce the approximately 32-day modulation of SN~2018bsz. Any successful explanation should account for its nearly constant characteristic timescale over the three observed cycles, the phase coherence from the optical to the NIR, the increasing amplitude towards shorter wavelengths, and the independent evidence that ejecta--CSM interaction occurred in this event. The present data do not uniquely identify a single mechanism.

\subsubsection{Lense--Thirring precession}
\label{sec:lt}

Lense--Thirring (LT) precession \citep{1975ApJ...195L..65B,1984GReGr..16..711M} of a fallback disk around a newborn magnetar has recently been proposed as an explanation for post-peak undulations in SLSNe-I, most notably for SN~2024afav \citep{2026Natur.651..321F}. In the specific model applied to SN~2024afav, the characteristic disk radius decreases as the disk falls towards the magnetar. Because the LT precession frequency scales as $\Omega_{\rm LT}\propto r^{-3}$, this evolution produces a chirped light curve in which successive modulations become progressively more frequent.

SN~2018bsz does not show the pronounced period decrease observed in SN~2024afav, making the rapidly infalling-disk realization inferred for that event less applicable. However, the absence of a detectable chirp does not exclude LT precession. A disk that remains near an equilibrium radius, or whose characteristic radius evolves slowly, may produce an approximately constant precession period. The same framework can also accommodate an approximately constant precession period if the disk remains near its equilibrium radius \citep{2026Natur.651..321F}. With only about three cycles, the observations of SN~2018bsz have limited sensitivity to moderate period evolution. LT precession therefore remains possible, although it is not uniquely required by the data.

\subsubsection{Structured CSM and pre-SN interaction}
\label{sec:pre_sn}

In a pre-SN interaction scenario, recurrent mass-loss episodes produce radially structured CSM before the explosion. The expanding SN ejecta then encounter successive density enhancements, converting kinetic energy into radiation. To first order, the observed interval between interaction episodes is

\begin{equation}
P_{\rm obs}\simeq\frac{\Delta R_{\rm CSM}}{V_{\rm sh}},
\end{equation}

where $\Delta R_{\rm CSM}$ is the radial separation between adjacent CSM structures and $V_{\rm sh}$ is the shock or ejecta velocity. If the structures were produced at intervals $P_{\rm pre}$ by material moving at velocity $V_{\rm w}$, then

\begin{equation}
P_{\rm pre}\simeq P_{\rm obs}\frac{V_{\rm sh}}{V_{\rm w}}.
\end{equation}

For $P_{\rm obs}\simeq32$\,days and $V_{\rm sh}\simeq10^{4}\,{\rm km\,s^{-1}}$, a wind velocity of $V_{\rm w}=100$--$1000\,{\rm km\,s^{-1}}$ corresponds to a pre-SN recurrence timescale of approximately 0.9--9\,yr. The inferred timescale is therefore of order one year for a fast WR wind and close to a decade for slower hydrogen-rich material.

Several independent observations indicate that ejecta--CSM interaction played an important role in SN~2018bsz. The emergence of multi-component hydrogen emission and the contemporaneous polarization evolution are consistent with interaction with highly aspherical, possibly disk-like, hydrogen-rich CSM \citep{mikka2022A&A...666A..30P}. The relatively flat late-time X-ray light curve is also more readily explained by ejecta--CSM interaction than by direct magnetar emission \citep{2026ApJ..1005..212A}. The late-time IR excess is best explained by newly formed amorphous-carbon dust in the metal-rich SN ejecta, rather than by an IR echo from pre-existing CSM or ISM dust \citep{janet2021arXiv210907942C}. Although this dust is not itself direct evidence for structured CSM, interaction-driven compression and cooling may have facilitated its formation. These observations make interaction with structured CSM a plausible origin of the photometric modulation, although they do not demonstrate that the modulation itself was interaction-powered.

The required CSM structure need not have been produced exclusively by binary interaction. Recurrent eruptive mass loss from a single star can also occur on year-to-decade timescales. The yellow hypergiant $\rho$~Cas, for example, has undergone mass-loss episodes separated by approximately 10--40\,yr, although the intervals are irregular \citep{2019MNRAS.483.3792K,2025A&A...694A.136V}. Pulsational pair-instability models can likewise produce multiple mass-ejection episodes separated by years to decades \citep{2017ApJ...836..244W}. However, these mechanisms do not necessarily produce regularly spaced structures, and their applicability to the stripped progenitor of an SLSN-I remains uncertain.

Pre-SN binary interaction is therefore an attractive possibility because orbital motion supplies a natural clock. The presence of hydrogen-rich CSM around a hydrogen-poor progenitor may be compatible with mass transfer or wind interaction in a WR+O binary. WR~140 provides a relevant Galactic analogue: it is a highly eccentric WC+O binary with an orbital period of approximately eight years whose recurrent dust-formation episodes have produced nested circumstellar dust shells \citep{wr140_2022NatAs...6.1308L}.

The approximately sinusoidal residuals do not uniquely constrain the orbital eccentricity. An eccentric binary may produce recurrent, pericentre-enhanced mass-loss episodes \citep{keiichi2026PASJ...78L...1M}, but the present photometry does not provide a direct measurement of the orbital geometry. A highly eccentric orbit therefore remains possible, but is neither demonstrated nor required by the light curve.

\subsubsection{Post-SN binary interaction}
\label{sec:post_sn}

Another possibility is post-SN binary interaction, as proposed for SN~2022jli based on its 12.4-day optical modulation, periodic H$\alpha$ velocity shifts, and the detection of a $\gamma$-ray source temporally and positionally consistent with the SN \citep{ping2024Natur.625..253C}. A subsequent analysis of this $\gamma$-ray source reported a consistent 12.5-day modulation, providing further support for the presence of a newly formed compact-object binary \citep{2025arXiv251209223Z}.
In this scenario, material transferred from a companion that survives the explosion is accreted by the newly formed compact remnant, which may be a neutron star (potentially a magnetar) or a stellar-mass black hole. The post-SN orbital motion then sets the recurrence timescale of the resulting energy injection. Binary population-synthesis calculations predict post-SN companion--compact-object interactions with orbital periods ranging from a few days to a few years \citep{andrea2026arXiv260725837E}. The approximately 32-day timescale of SN~2018bsz lies within this predicted range, although the fraction of such systems that produce detectable light-curve modulations remains uncertain.

A related, although more speculative, configuration could arise if a natal kick tilted the surviving post-SN orbit relative to pre-existing equatorial CSM. The compact remnant could then pass through the CSM plane at recurrent orbital phases and produce additional energy injection. Depending on the geometry, one or two interaction episodes could occur per orbit. The present observations provide no direct constraint on such a configuration, but it represents another possible connection between the post-SN orbit and the pre-existing disk-like CSM.

Accretion onto the compact object could, in principle, contribute substantially to the energy budget of SN~2018bsz. However, neither the required accreted mass nor the efficiency with which accretion energy is converted into observable optical radiation is constrained by the present observations, and quantitative modelling is required \citep{hirai2025ApJ...995...55H}. A further challenge is whether post-SN accretion can simultaneously reproduce the multi-component hydrogen emission, polarization evolution, and other observables attributed to ejecta--CSM interaction. These properties arise more directly in the pre-SN structured-CSM scenario \citep{mikka2022A&A...666A..30P}.

\section{Conclusion} \label{sec:conclusion}

We have presented a multiband photometric reanalysis of the nearby hydrogen-poor superluminous SN~2018bsz using \textit{Swift}/UVOT and GROND observations spanning approximately 110 rest-frame days after maximum light. After modelling and subtracting the smooth underlying light-curve evolution, we identify a coherent, quasi-periodic modulation across ten bands from $u$ to $K_s$. The most significant individual-band detections, in the $g$, $r$, $i$, $z$, and $J$ bands, yield best-fitting periods of approximately 31.5--31.8 days. The joint multiband analysis gives a rest-frame period of $P=31.61^{+0.03}_{-0.03}$ days.

The modulation can be traced over approximately three cycles, with its peaks and troughs occurring at broadly consistent phases across wavelength. However, its amplitude increases towards shorter wavelengths. The lower-significance bands also show broader and somewhat shifted periodogram peaks, ranging from approximately 30 to 35 days, which may indicate a wavelength dependence of the best-fitting recurrence timescale. The signal is therefore broadly phase-coherent across wavelength, but is not strictly achromatic in either its amplitude or possibly its characteristic timescale.

The recovered period and phase structure remain consistent when alternative descriptions of the underlying light-curve evolution are adopted. Reference stars measured in the same GROND images show no corresponding coherent variation, and the difference between the recovered period and the lunar synodic cycle produces a progressive phase offset over the observing interval. Variations at consistent phases in the space-based \textit{Swift}/UVOT photometry provide an additional check against atmospheric or ground-based instrumental systematics. Taken together, these tests disfavour observational systematics as the origin of the modulation.

The primary result of this work is the identification and characterization of the recurrent modulation rather than the determination of its physical origin. Its lack of pronounced period evolution makes an SN~2024afav-like, rapidly chirping realization of LT precession less compelling, although the limited number of observed cycles does not exclude LT precession more generally \citep{2026Natur.651..321F}. Interaction with CSM structured before the explosion provides another possible interpretation, particularly given the independent evidence for ejecta--CSM interaction in SN~2018bsz \citep{mikka2022A&A...666A..30P,2026ApJ..1005..212A}. Alternatively, post-SN accretion from a surviving companion onto the compact remnant could produce orbitally modulated energy injection, as proposed for SN~2022jli \citep{ping2024Natur.625..253C,hirai2025ApJ...995...55H}. The present observations do not uniquely distinguish among these possibilities, which need not be mutually exclusive.

SN~2018bsz therefore provides one of the clearest examples to date of a coherent, month-scale modulation across the optical and NIR light curves of an SLSN-I. The wavelength dependence of the modulation amplitude, together with the possible wavelength dependence of its best-fitting recurrence timescale, provides an additional observational constraint on its physical origin. Future models should reproduce not only the overall recurrence timescale and phase coherence, but also these wavelength-dependent properties. Distinguishing among the possible mechanisms will require similarly well-sampled events covering more cycles, accompanied by densely sampled, phase-resolved spectroscopy.

\section*{Acknowledgments}
 A.S.K., T.-W.C. and A.A. acknowledge the funding support from the Yushan Young Fellow Program by the Ministry of Education, Taiwan (MOE-111-YSFMS-0008-001-P1), and the National Science and Technology Council, Taiwan (NSTC grant 114-2112-M-008-021-MY3).
 K.M. acknowledges support from JSPS KAKENHI grant (JP24KK0070, JP24H01810, and JP23H04894). 
 This publication has emanated from research conducted with the financial
support of Taighde \'{E}ireann--Research Ireland under Grant No.~24/FPP-P/12959.

During the preparation of this manuscript, the authors used ChatGPT (OpenAI; accessed 2026 September) to assist with English-language editing and to improve the clarity and organization of the text. ChatGPT was not used to generate or analyse the observational data. All AI-assisted suggestions were critically evaluated and revised by the authors, and all scientific statements and references were independently verified against the original sources. The authors take full responsibility for the scientific content, conclusions, and final text of the manuscript.

\software{ChatGPT (OpenAI), AutoPhOT \citep{autophot2022A&A...667A..62B}, GLS \citep{astropy2022ApJ...935..167A}}

\restartappendixnumbering
\appendix

\section{Alternative polynomial detrending}
\label{app:detrending}

\begin{figure}[ht!]
    \centering
    \includegraphics[width=0.75\textwidth]{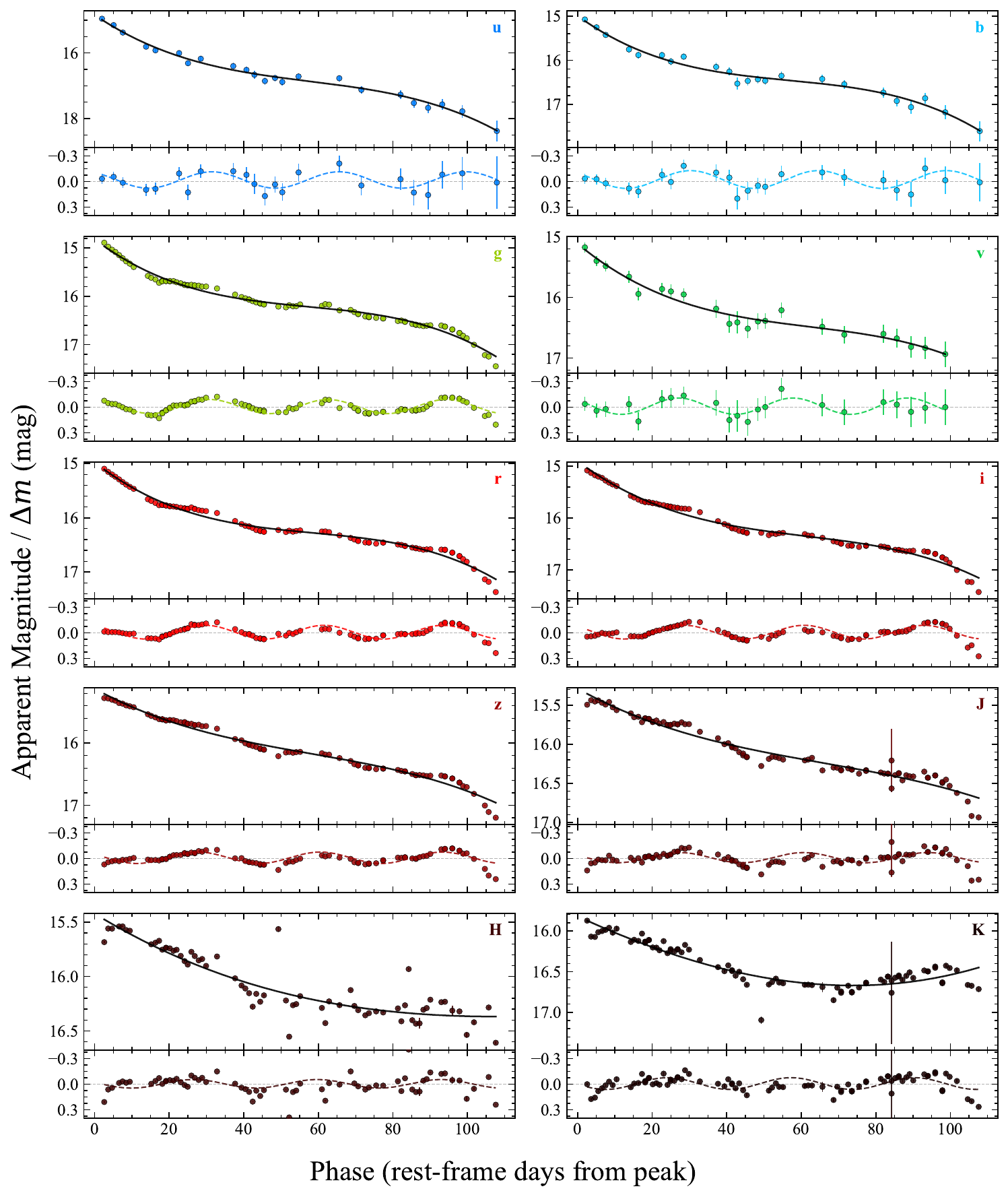}
    \caption{Residuals obtained after detrending with a
    third-degree polynomial, following the same procedure as in
    Figure~\ref{fig:detrending}.}
    \label{fig:detrend3}
\end{figure}

\begin{figure}[ht!]
    \centering
    \includegraphics[width=0.75\textwidth]{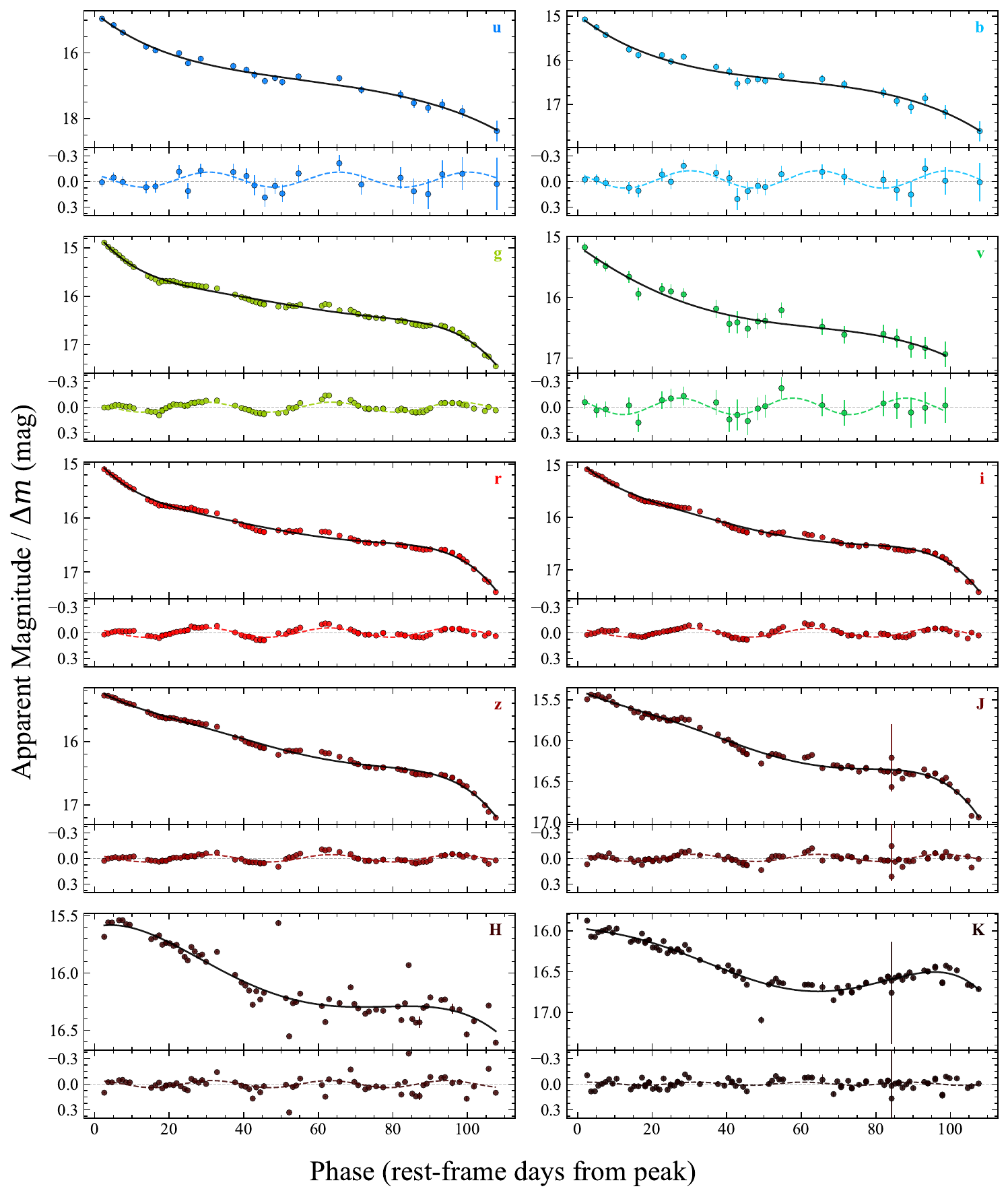}
    \caption{Residuals obtained after detrending with a
    fifth-degree polynomial, following the same procedure as in
    Figure~\ref{fig:detrending}.}
    \label{fig:detrend5}
\end{figure}

\begin{figure}[ht!] \centering \includegraphics[width=0.75\textwidth]{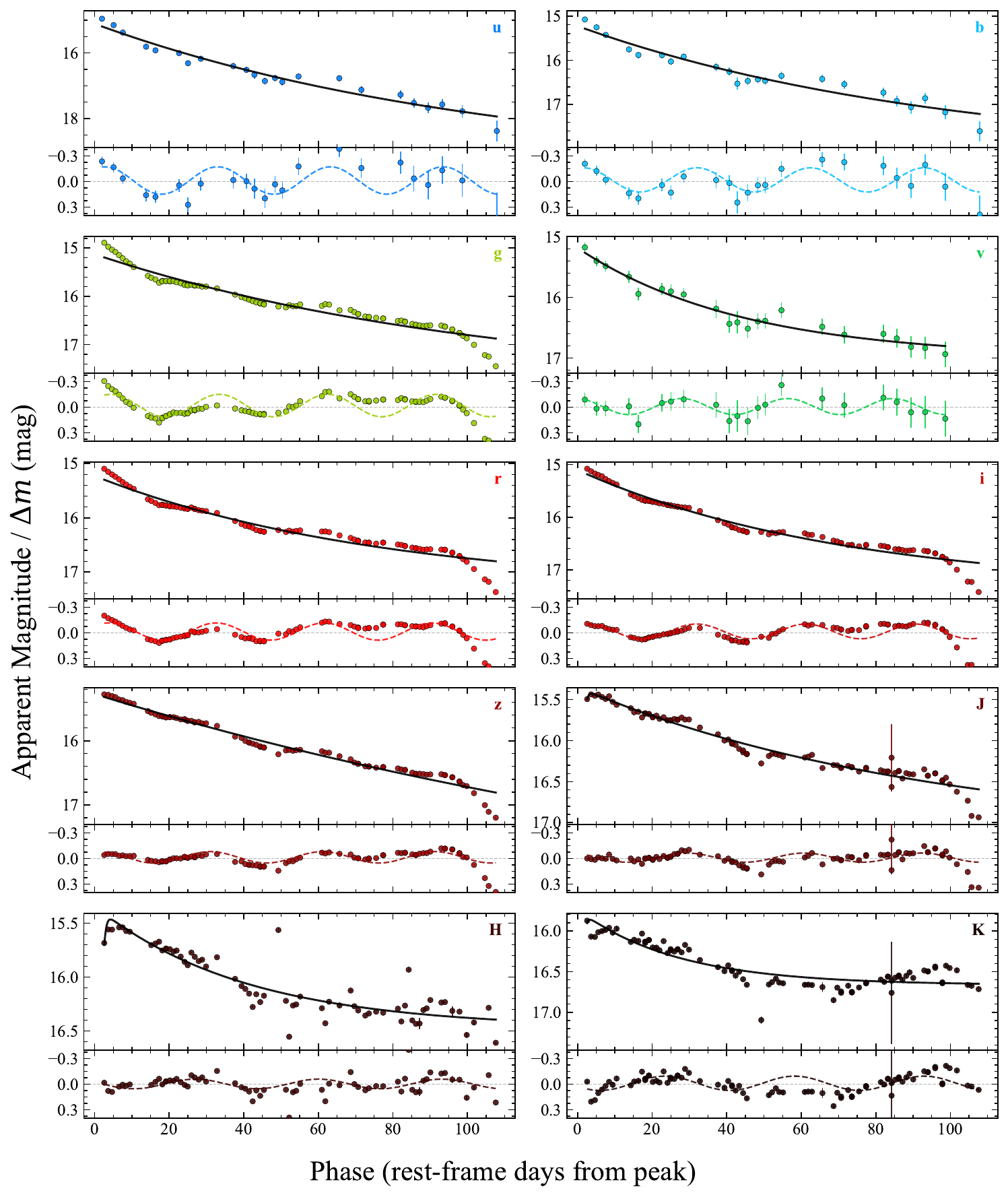} \caption{Residuals obtained after modelling and subtracting the underlying light-curve evolution with a Bazin function, instead of the fourth-degree polynomial adopted in the main analysis. All other procedures and symbols are the same as in Figure~\ref{fig:detrending}.} \label{fig:detrend_bazin} \end{figure}

\bibliographystyle{aasjournalv7.1}
\bibliography{citation}

\begin{thebibliography}{}
\expandafter\ifx\csname natexlab\endcsname\relax\def\natexlab#1{#1}\fi
\providecommand{\url}[1]{\href{#1}{#1}}
\providecommand{\dodoi}[1]{doi:~\href{http://doi.org/#1}{\nolinkurl{#1}}}
\providecommand{\doeprint}[1]{\href{http://ascl.net/#1}{\nolinkurl{http://ascl.net/#1}}}
\providecommand{\doarXiv}[1]{\href{https://arxiv.org/abs/#1}{\nolinkurl{https://arxiv.org/abs/#1}}}

\bibitem[{J. {Ahlvind} {et~al.}(2026){Ahlvind}, {Larsson}, {Alp}, \& {Lunnan}}]{2026ApJ..1005..212A}
{Ahlvind}, J., {Larsson}, J., {Alp}, D., \& {Lunnan}, R. 2026, \bibinfo{title}{{Multiepoch X-Ray Detection of SLSN-I 2018bsz: Constraints on the Powering Mechanism and Ejecta Structure},} \apj, 1005, 212, \dodoi{10.3847/1538-4357/ae7c7a}

\bibitem[{J.~P. {Anderson} {et~al.}(2018){Anderson}, {Pessi}, {Dessart}, {Inserra}, {Hiramatsu}, {Taggart}, {Smartt}, {Leloudas}, {Chen}, {M{\"o}ller}, {Roy}, {Schulze}, {Perley}, {Selsing}, {Prentice}, {Gal-Yam}, {Angus}, {Arcavi}, {Ashall}, {Bulla}, {Bray}, {Burke}, {Callis}, {Cartier}, {Chang}, {Chambers}, {Clark}, {Denneau}, {Dennefeld}, {Flewelling}, {Fraser}, {Galbany}, {Gromadzki}, {Guti{\'e}rrez}, {Heinze}, {Hosseinzadeh}, {Howell}, {Hsiao}, {Kankare}, {Kostrzewa-Rutkowska}, {Magnier}, {Maguire}, {Mazzali}, {McBrien}, {McCully}, {Morrell}, {Lowe}, {Onken}, {Onori}, {Phillips}, {Rest}, {Ridden-Harper}, {Ruiter}, {Sand}, {Smith}, {Smith}, {Stalder}, {Stritzinger}, {Sullivan}, {Tonry}, {Tucker}, {Valenti}, {Wainscoat}, {Waters}, {Wolf}, \& {Young}}]{anderson2018bsz2018A&A...620A..67A}
{Anderson}, J.~P., {Pessi}, P.~J., {Dessart}, L., {et~al.} 2018, \bibinfo{title}{{A nearby super-luminous supernova with a long pre-maximum \& ``plateau'' and strong C II features},} \aap, 620, A67, \dodoi{10.1051/0004-6361/201833725}

\bibitem[{{}{Astropy Collaboration} {et~al.}(2022){Astropy Collaboration}, {Price-Whelan}, {Lim}, {Earl}, {Starkman}, {Bradley}, {Shupe}, {Patil}, {Corrales}, {Brasseur}, {N{\"o}the}, {Donath}, {Tollerud}, {Morris}, {Ginsburg}, {Vaher}, {Weaver}, {Tocknell}, {Jamieson}, {van Kerkwijk}, {Robitaille}, {Merry}, {Bachetti}, {G{\"u}nther}, {Aldcroft}, {Alvarado-Montes}, {Archibald}, {B{\'o}di}, {Bapat}, {Barentsen}, {Baz{\'a}n}, {Biswas}, {Boquien}, {Burke}, {Cara}, {Cara}, {Conroy}, {Conseil}, {Craig}, {Cross}, {Cruz}, {D'Eugenio}, {Dencheva}, {Devillepoix}, {Dietrich}, {Eigenbrot}, {Erben}, {Ferreira}, {Foreman-Mackey}, {Fox}, {Freij}, {Garg}, {Geda}, {Glattly}, {Gondhalekar}, {Gordon}, {Grant}, {Greenfield}, {Groener}, {Guest}, {Gurovich}, {Handberg}, {Hart}, {Hatfield-Dodds}, {Homeier}, {Hosseinzadeh}, {Jenness}, {Jones}, {Joseph}, {Kalmbach}, {Karamehmetoglu}, {Ka{\l}uszy{\'n}ski}, {Kelley}, {Kern}, {Kerzendorf}, {Koch}, {Kulumani}, {Lee}, {Ly}, {Ma}, {MacBride}, {Maljaars}, {Muna}, {Murphy}, {Norman},
  {O'Steen}, {Oman}, {Pacifici}, {Pascual}, {Pascual-Granado}, {Patil}, {Perren}, {Pickering}, {Rastogi}, {Roulston}, {Ryan}, {Rykoff}, {Sabater}, {Sakurikar}, {Salgado}, {Sanghi}, {Saunders}, {Savchenko}, {Schwardt}, {Seifert-Eckert}, {Shih}, {Jain}, {Shukla}, {Sick}, {Simpson}, {Singanamalla}, {Singer}, {Singhal}, {Sinha}, {Sip{\H{o}}cz}, {Spitler}, {Stansby}, {Streicher}, {{\v{S}}umak}, {Swinbank}, {Taranu}, {Tewary}, {Tremblay}, {de Val-Borro}, {Van Kooten}, {Vasovi{\'c}}, {Verma}, {de Miranda Cardoso}, {Williams}, {Wilson}, {Winkel}, {Wood-Vasey}, {Xue}, {Yoachim}, {Zhang}, {Zonca}, \& {Astropy Project Contributors}}]{astropy2022ApJ...935..167A}
{Astropy Collaboration}, {Price-Whelan}, A.~M., {Lim}, P.~L., {et~al.} 2022, \bibinfo{title}{{The Astropy Project: Sustaining and Growing a Community-oriented Open-source Project and the Latest Major Release (v5.0) of the Core Package},} \apj, 935, 167, \dodoi{10.3847/1538-4357/ac7c74}

\bibitem[{R.~V. {Baluev}(2008){Baluev}}]{fap2008MNRAS.385.1279B}
{Baluev}, R.~V. 2008, \bibinfo{title}{{Assessing the statistical significance of periodogram peaks},} \mnras, 385, 1279, \dodoi{10.1111/j.1365-2966.2008.12689.x}

\bibitem[{J.~M. {Bardeen} {\&} J.~A. {Petterson}(1975){Bardeen} \& {Petterson}}]{1975ApJ...195L..65B}
{Bardeen}, J.~M., \& {Petterson}, J.~A. 1975, \bibinfo{title}{{The Lense-Thirring Effect and Accretion Disks around Kerr Black Holes},} \apjl, 195, L65, \dodoi{10.1086/181711}

\bibitem[{S.~J. {Brennan} {\&} M. {Fraser}(2022){Brennan} \& {Fraser}}]{autophot2022A&A...667A..62B}
{Brennan}, S.~J., \& {Fraser}, M. 2022, \bibinfo{title}{{The Automated Photometry of Transients pipeline (AUTOPHOT)},} \aap, 667, A62, \dodoi{10.1051/0004-6361/202243067}

\bibitem[{R. {Cartier} {et~al.}(2026){Cartier}, {Contreras}, {Stritzinger}, {Hamuy}, {Ruiz-Lapuente}, {Prieto}, {Anderson}, {Cikota}, \& {Gerlach}}]{2026A&A...707A.161C}
{Cartier}, R., {Contreras}, C., {Stritzinger}, M., {et~al.} 2026, \bibinfo{title}{{Unveiling the nature of SN 2022jli: The first double-peaked stripped-envelope supernova showing periodic undulations and dust emission at late times},} \aap, 707, A161, \dodoi{10.1051/0004-6361/202452729}

\bibitem[{P. {Chen} {et~al.}(2024){Chen}, {Gal-Yam}, {Sollerman}, {Schulze}, {Post}, {Liu}, {Ofek}, {Das}, {Fremling}, {Horesh}, {Katz}, {Kushnir}, {Kasliwal}, {Kulkarni}, {Liu}, {Liu}, {Miller}, {Rose}, {Waxman}, {Yang}, {Yao}, {Zackay}, {Bellm}, {Dekany}, {Drake}, {Fang}, {Fynbo}, {Groom}, {Helou}, {Irani}, {Jegou du Laz}, {Liu}, {Mazzali}, {Neill}, {Qin}, {Riddle}, {Sharon}, {Strotjohann}, {Wold}, \& {Yan}}]{ping2024Natur.625..253C}
{Chen}, P., {Gal-Yam}, A., {Sollerman}, J., {et~al.} 2024, \bibinfo{title}{{A 12.4-day periodicity in a close binary system after a supernova},} \nat, 625, 253, \dodoi{10.1038/s41586-023-06787-x}

\bibitem[{T.-W. {Chen} {et~al.}(2018){Chen}, {Inserra}, {Fraser}, {Moriya}, {Schady}, {Schweyer}, {Filippenko}, {Perley}, {Ruiter}, {Seitenzahl}, {Sollerman}, {Taddia}, {Anderson}, {Foley}, {Jerkstrand}, {Ngeow}, {Pan}, {Pastorello}, {Points}, {Smartt}, {Smith}, {Taubenberger}, {Wiseman}, {Young}, {Benetti}, {Berton}, {Bufano}, {Clark}, {Della Valle}, {Galbany}, {Gal-Yam}, {Gromadzki}, {Guti{\'e}rrez}, {Heinze}, {Kankare}, {Kilpatrick}, {Kuncarayakti}, {Leloudas}, {Lin}, {Maguire}, {Mazzali}, {McBrien}, {Prentice}, {Rau}, {Rest}, {Siebert}, {Stalder}, {Tonry}, \& {Yu}}]{2018ApJ...867L..31C}
{Chen}, T.-W., {Inserra}, C., {Fraser}, M., {et~al.} 2018, \bibinfo{title}{{SN 2017ens: The Metamorphosis of a Luminous Broadlined Type Ic Supernova into an SN IIn},} \apjl, 867, L31, \dodoi{10.3847/2041-8213/aaeb2e}

\bibitem[{T.-W. {Chen} {et~al.}(2021){Chen}, {Brennan}, {Wesson}, {Fraser}, {Schweyer}, {Inserra}, {Schulze}, {Nicholl}, {Anderson}, {Hsiao}, {Jerkstrand}, {Kankare}, {Kool}, {Kravtsov}, {Kuncarayakti}, {Leloudas}, {Li}, {Matsuura}, {Pursiainen}, {Roy}, {Ruiter}, {Schady}, {Seitenzahl}, {Sollerman}, {Tartaglia}, {Wang}, {Yates}, {Yang}, {Baade}, {Carini}, {Gal-Yam}, {Galbany}, {Gonzalez-Gaitan}, {Gromadzki}, {Gutierrez}, {Kotak}, {Maguire}, {Mazzali}, {Mueller-Bravo}, {Paraskeva}, {Pessi}, {Pignata}, {Rau}, \& {Young}}]{janet2021arXiv210907942C}
{Chen}, T.-W., {Brennan}, S.~J., {Wesson}, R., {et~al.} 2021, \bibinfo{title}{{SN 2018bsz: significant dust formation in a nearby superluminous supernova},} arXiv e-prints, arXiv:2109.07942, \dodoi{10.48550/arXiv.2109.07942}

\bibitem[{Z.~H. {Chen} {et~al.}(2023{\natexlab{a}}){Chen}, {Yan}, {Kangas}, {Lunnan}, {Sollerman}, {Schulze}, {Perley}, {Chen}, {Taggart}, {Hinds}, {Gal-Yam}, {Wang}, {De}, {Bellm}, {Bloom}, {Dekany}, {Graham}, {Kasliwal}, {Kulkarni}, {Laher}, {Neill}, \& {Rusholme}}]{2023ApJ...943...42C}
{Chen}, Z.~H., {Yan}, L., {Kangas}, T., {et~al.} 2023{\natexlab{a}}, \bibinfo{title}{{The Hydrogen-poor Superluminous Supernovae from the Zwicky Transient Facility Phase I Survey. II. Light-curve Modeling and Characterization of Undulations},} \apj, 943, 42, \dodoi{10.3847/1538-4357/aca162}

\bibitem[{A. {Ercolino}(2026){Ercolino}}]{andrea2026arXiv260725837E}
{Ercolino}, A. 2026, \bibinfo{title}{{Progenitor models of supernovae interacting with their binary companions},} arXiv e-prints, arXiv:2607.25837, \dodoi{10.48550/arXiv.2607.25837}

\bibitem[{J.~R. {Farah} {et~al.}(2026){Farah}, {Prust}, {Howell}, {Ni}, {McCully}, {Andrews}, {Kumar}, {Hiramatsu}, {Gomez}, {Wynn}, {Filippenko}, {Bostroem}, {Berger}, \& {Blanchard}}]{2026Natur.651..321F}
{Farah}, J.~R., {Prust}, L.~J., {Howell}, D.~A., {et~al.} 2026, \bibinfo{title}{{Lense─Thirring precessing magnetar engine drives a superluminous supernova},} \nat, 651, 321, \dodoi{10.1038/s41586-026-10151-0}

\bibitem[{A. {Fiore} {et~al.}(2021){Fiore}, {Chen}, {Jerkstrand}, {Benetti}, {Ciolfi}, {Inserra}, {Cappellaro}, {Pastorello}, {Leloudas}, {Schulze}, {Berton}, {Burke}, {McCully}, {Fong}, {Galbany}, {Gromadzki}, {Guti{\'e}rrez}, {Hiramatsu}, {Hosseinzadeh}, {Howell}, {Kankare}, {Lunnan}, {M{\"u}ller-Bravo}, {O'Neill}, {Nicholl}, {Rau}, {Sollerman}, {Terreran}, {Valenti}, \& {Young}}]{2021MNRAS.502.2120F}
{Fiore}, A., {Chen}, T.-W., {Jerkstrand}, A., {et~al.} 2021, \bibinfo{title}{{SN 2017gci: a nearby Type I Superluminous Supernova with a bumpy tail},} \mnras, 502, 2120, \dodoi{10.1093/mnras/staa4035}

\bibitem[{A. {Gal-Yam}(2019){Gal-Yam}}]{galyam2019ARA&A..57..305G}
{Gal-Yam}, A. 2019, \bibinfo{title}{{The Most Luminous Supernovae},} \araa, 57, 305, \dodoi{10.1146/annurev-astro-081817-051819}

\bibitem[{N. {Gehrels}(1986){Gehrels}}]{Gehrels1986ApJ...303..336G}
{Gehrels}, N. 1986, \bibinfo{title}{{Confidence Limits for Small Numbers of Events in Astrophysical Data},} \apj, 303, 336, \dodoi{10.1086/164079}

\bibitem[{J. {Greiner} {et~al.}(2008){Greiner}, {Bornemann}, {Clemens}, {Deuter}, {Hasinger}, {Honsberg}, {Huber}, {Huber}, {Krauss}, {Kr{\"u}hler}, {K{\"u}pc{\"u} Yolda{\textcommabelow s}}, {Mayer-Hasselwander}, {Mican}, {Primak}, {Schrey}, {Steiner}, {Szokoly}, {Th{\"o}ne}, {Yolda{\textcommabelow s}}, {Klose}, {Laux}, \& {Winkler}}]{GROND2008PASP..120..405G}
{Greiner}, J., {Bornemann}, W., {Clemens}, C., {et~al.} 2008, \bibinfo{title}{{GROND{\textemdash}a 7-Channel Imager},} \pasp, 120, 405, \dodoi{10.1086/587032}

\bibitem[{R. {Hirai} {et~al.}(2025){Hirai}, {Podsiadlowski}, {Hoeflich}, {Barkov}, {Chan}, {Liptai}, \& {Nagataki}}]{hirai2025ApJ...995...55H}
{Hirai}, R., {Podsiadlowski}, P., {Hoeflich}, P., {et~al.} 2025, \bibinfo{title}{{Supernova-induced Binary-interaction-powered Supernovae: A Model for SN2022jli},} \apj, 995, 55, \dodoi{10.3847/1538-4357/ae172e}

\bibitem[{A. {Horowicz} {et~al.}(2026){Horowicz}, {Ofek}, \& {Gal-Yam}}]{Horowicz2026arXiv260818207H}
{Horowicz}, A., {Ofek}, E.~O., \& {Gal-Yam}, A. 2026, \bibinfo{title}{{The fraction of periodic SN Ib/c light curves},} arXiv e-prints, arXiv:2608.18207, \dodoi{10.48550/arXiv.2608.18207}

\bibitem[{G. {Hosseinzadeh} {et~al.}(2022){Hosseinzadeh}, {Berger}, {Metzger}, {Gomez}, {Nicholl}, \& {Blanchard}}]{Hosseinzadeh2022ApJ...933...14H}
{Hosseinzadeh}, G., {Berger}, E., {Metzger}, B.~D., {et~al.} 2022, \bibinfo{title}{{Bumpy Declining Light Curves Are Common in Hydrogen-poor Superluminous Supernovae},} \apj, 933, 14, \dodoi{10.3847/1538-4357/ac67dd}

\bibitem[{C. {Inserra} {et~al.}(2017){Inserra}, {Nicholl}, {Chen}, {Jerkstrand}, {Smartt}, {Kr{\"u}hler}, {Anderson}, {Baltay}, {Della Valle}, {Fraser}, {Gal-Yam}, {Galbany}, {Kankare}, {Maguire}, {Rabinowitz}, {Smith}, {Valenti}, \& {Young}}]{2017MNRAS.468.4642I}
{Inserra}, C., {Nicholl}, M., {Chen}, T.-W., {et~al.} 2017, \bibinfo{title}{{Complexity in the light curves and spectra of slow-evolving superluminous supernovae},} \mnras, 468, 4642, \dodoi{10.1093/mnras/stx834}

\bibitem[{A. {Jones} {et~al.}(2013){Jones}, {Noll}, {Kausch}, {Szyszka}, \& {Kimeswenger}}]{Jones2013A&A...560A..91J}
{Jones}, A., {Noll}, S., {Kausch}, W., {Szyszka}, C., \& {Kimeswenger}, S. 2013, \bibinfo{title}{{An advanced scattered moonlight model for Cerro Paranal},} \aap, 560, A91, \dodoi{10.1051/0004-6361/201322433}

\bibitem[{M. {Kraus} {et~al.}(2019){Kraus}, {Kolka}, {Aret}, {Nickeler}, {Maravelias}, {Eenm{\"a}e}, {Lobel}, \& {Klochkova}}]{2019MNRAS.483.3792K}
{Kraus}, M., {Kolka}, I., {Aret}, A., {et~al.} 2019, \bibinfo{title}{{A new outburst of the yellow hypergiant star {\ensuremath{\rho}} Cas},} \mnras, 483, 3792, \dodoi{10.1093/mnras/sty3375}

\bibitem[{K. {Krisciunas} {\&} B.~E. {Schaefer}(1991){Krisciunas} \& {Schaefer}}]{Krisciunas1991PASP..103.1033K}
{Krisciunas}, K., \& {Schaefer}, B.~E. 1991, \bibinfo{title}{{A Model of the Brightness of Moonlight},} \pasp, 103, 1033, \dodoi{10.1086/132921}

\bibitem[{H. {Kumar} {et~al.}(2026){Kumar}, {Blanchard}, {Berger}, {Athukoralalage}, {Hiramatsu}, {Gomez}, {Andrews}, {Bostroem}, {Farah}, {Howell}, \& {McCully}}]{kumar2026ApJ...998L...3K}
{Kumar}, H., {Blanchard}, P.~K., {Berger}, E., {et~al.} 2026, \bibinfo{title}{{SN 2024afav: A Superluminous Supernova with Multiple Light-curve Bumps and Spectroscopic Signatures of Circumstellar Interaction},} \apjl, 998, L3, \dodoi{10.3847/2041-8213/ae3749}

\bibitem[{R.~M. {Lau} {et~al.}(2022){Lau}, {Hankins}, {Han}, {Argyriou}, {Corcoran}, {Eldridge}, {Endo}, {Fox}, {Garcia Marin}, {Gull}, {Jones}, {Hamaguchi}, {Lamberts}, {Law}, {Madura}, {Marchenko}, {Matsuhara}, {Moffat}, {Morris}, {Morris}, {Onaka}, {Ressler}, {Richardson}, {Russell}, {Sanchez-Bermudez}, {Smith}, {Soulain}, {Stevens}, {Tuthill}, {Weigelt}, {Williams}, \& {Yamaguchi}}]{wr140_2022NatAs...6.1308L}
{Lau}, R.~M., {Hankins}, M.~J., {Han}, Y., {et~al.} 2022, \bibinfo{title}{{Nested dust shells around the Wolf-Rayet binary WR 140 observed with JWST},} Nature Astronomy, 6, 1308, \dodoi{10.1038/s41550-022-01812-x}

\bibitem[{K. {Maeda} {et~al.}(2026){Maeda}, {Kuncarayakti}, {Nagao}, {Kawabata}, {Taguchi}, {Uno}, \& {De}}]{keiichi2026PASJ...78L...1M}
{Maeda}, K., {Kuncarayakti}, H., {Nagao}, T., {et~al.} 2026, \bibinfo{title}{{Peculiar SN Ic 2022esa: An explosion of a massive Wolf─Rayet star in a binary as a precursor to a BH─BH binary?},} \pasj, 78, L1, \dodoi{10.1093/pasj/psaf140}

\bibitem[{B. {Mashhoon} {et~al.}(1984){Mashhoon}, {Hehl}, \& {Theiss}}]{1984GReGr..16..711M}
{Mashhoon}, B., {Hehl}, F.~W., \& {Theiss}, D.~S. 1984, \bibinfo{title}{{On the gravitational effects of rotating masses: the Thirring-Lense papers.},} General Relativity and Gravitation, 16, 711, \dodoi{10.1007/BF00762913}

\bibitem[{T. {Moore} {et~al.}(2023){Moore}, {Smartt}, {Nicholl}, {Srivastav}, {Stevance}, {Jess}, {Grant}, {Fulton}, {Rhodes}, {Sim}, {Hirai}, {Podsiadlowski}, {Anderson}, {Ashall}, {Bate}, {Fender}, {Guti{\'e}rrez}, {Howell}, {Huber}, {Inserra}, {Leloudas}, {Monard}, {M{\"u}ller-Bravo}, {Shappee}, {Smith}, {Terreran}, {Tonry}, {Tucker}, {Young}, {Aamer}, {Chen}, {Ragosta}, {Galbany}, {Gromadzki}, {Harvey}, {Hoeflich}, {McCully}, {Newsome}, {Gonzalez}, {Pellegrino}, {Ramsden}, {P{\'e}rez-Torres}, {Ridley}, {Sheng}, \& {Weston}}]{thomas2023ApJ...956L..31M}
{Moore}, T., {Smartt}, S.~J., {Nicholl}, M., {et~al.} 2023, \bibinfo{title}{{SN 2022jli: A Type Ic Supernova with Periodic Modulation of Its Light Curve and an Unusually Long Rise},} \apjl, 956, L31, \dodoi{10.3847/2041-8213/acfc25}

\bibitem[{T.~J. {Moriya} {et~al.}(2018){Moriya}, {Sorokina}, \& {Chevalier}}]{moriya2018SSRv..214...59M}
{Moriya}, T.~J., {Sorokina}, E.~I., \& {Chevalier}, R.~A. 2018, \bibinfo{title}{{Superluminous Supernovae},} \ssr, 214, 59, \dodoi{10.1007/s11214-018-0493-6}

\bibitem[{M. {Nicholl} {et~al.}(2016){Nicholl}, {Berger}, {Smartt}, {Margutti}, {Kamble}, {Alexander}, {Chen}, {Inserra}, {Arcavi}, {Blanchard}, {Cartier}, {Chambers}, {Childress}, {Chornock}, {Cowperthwaite}, {Drout}, {Flewelling}, {Fraser}, {Gal-Yam}, {Galbany}, {Harmanen}, {Holoien}, {Hosseinzadeh}, {Howell}, {Huber}, {Jerkstrand}, {Kankare}, {Kochanek}, {Lin}, {Lunnan}, {Magnier}, {Maguire}, {McCully}, {McDonald}, {Metzger}, {Milisavljevic}, {Mitra}, {Reynolds}, {Saario}, {Shappee}, {Smith}, {Valenti}, {Villar}, {Waters}, \& {Young}}]{2015bn2016ApJ...826...39N}
{Nicholl}, M., {Berger}, E., {Smartt}, S.~J., {et~al.} 2016, \bibinfo{title}{{SN 2015BN: A Detailed Multi-wavelength View of a Nearby Superluminous Supernova},} \apj, 826, 39, \dodoi{10.3847/0004-637X/826/1/39}

\bibitem[{M. {Pursiainen} {et~al.}(2022){Pursiainen}, {Leloudas}, {Paraskeva}, {Cikota}, {Anderson}, {Angus}, {Brennan}, {Bulla}, {Camacho-I{\~n}iguez}, {Charalampopoulos}, {Chen}, {Delgado Manche{\~n}o}, {Fraser}, {Frohmaier}, {Galbany}, {Guti{\'e}rrez}, {Gromadzki}, {Inserra}, {Maund}, {M{\"u}ller-Bravo}, {Mu{\~n}oz Torres}, {Nicholl}, {Onori}, {Patat}, {Pessi}, {Roy}, {Spyromilio}, {Wiseman}, \& {Young}}]{mikka2022A&A...666A..30P}
{Pursiainen}, M., {Leloudas}, G., {Paraskeva}, E., {et~al.} 2022, \bibinfo{title}{{SN 2018bsz: A Type I superluminous supernova with aspherical circumstellar material},} \aap, 666, A30, \dodoi{10.1051/0004-6361/202243256}

\bibitem[{R.~M. {Quimby} {et~al.}(2011){Quimby}, {Kulkarni}, {Kasliwal}, {Gal-Yam}, {Arcavi}, {Sullivan}, {Nugent}, {Thomas}, {Howell}, {Nakar}, {Bildsten}, {Theissen}, {Law}, {Dekany}, {Rahmer}, {Hale}, {Smith}, {Ofek}, {Zolkower}, {Velur}, {Walters}, {Henning}, {Bui}, {McKenna}, {Poznanski}, {Cenko}, \& {Levitan}}]{quimby2011Natur.474..487Q}
{Quimby}, R.~M., {Kulkarni}, S.~R., {Kasliwal}, M.~M., {et~al.} 2011, \bibinfo{title}{{Hydrogen-poor superluminous stellar explosions},} \nat, 474, 487, \dodoi{10.1038/nature10095}

\bibitem[{P.~W.~A. {Roming} {et~al.}(2005){Roming}, {Kennedy}, {Mason}, {Nousek}, {Ahr}, {Bingham}, {Broos}, {Carter}, {Hancock}, {Huckle}, {Hunsberger}, {Kawakami}, {Killough}, {Koch}, {McLelland}, {Smith}, {Smith}, {Soto}, {Boyd}, {Breeveld}, {Holland}, {Ivanushkina}, {Pryzby}, {Still}, \& {Stock}}]{uvot2005SSRv..120...95R}
{Roming}, P. W.~A., {Kennedy}, T.~E., {Mason}, K.~O., {et~al.} 2005, \bibinfo{title}{{The Swift Ultra-Violet/Optical Telescope},} \ssr, 120, 95, \dodoi{10.1007/s11214-005-5095-4}

\bibitem[{A.~M. {van Genderen} {et~al.}(2025){van Genderen}, {Lobel}, {Timmerman}, {Deul}, {Vos}, {Nieuwenhuijzen}, {van Ballegoij}, {Sblewski}, {Henry}, {Blown}, \& {Di Scala}}]{2025A&A...694A.136V}
{van Genderen}, A.~M., {Lobel}, A., {Timmerman}, R., {et~al.} 2025, \bibinfo{title}{{Investigation of the pulsations, outbursts, and evolution of the yellow hypergiants: {\ensuremath{\rho}} Cas, HR 8752, and HR 5171A, with notes on HD 179821},} \aap, 694, A136, \dodoi{10.1051/0004-6361/202449384}

\bibitem[{J.~T. {VanderPlas}(2018){VanderPlas}}]{lombscargle2018ApJS..236...16V}
{VanderPlas}, J.~T. 2018, \bibinfo{title}{{Understanding the Lomb-Scargle Periodogram},} \apjs, 236, 16, \dodoi{10.3847/1538-4365/aab766}

\bibitem[{J.~T. {VanderPlas} {\&} {\v{Z}}. {Ivezi{\'c}}(2015){VanderPlas} \& {Ivezi{\'c}}}]{multiband2015ApJ...812...18V}
{VanderPlas}, J.~T., \& {Ivezi{\'c}}, {\v{Z}}. 2015, \bibinfo{title}{{Periodograms for Multiband Astronomical Time Series},} \apj, 812, 18, \dodoi{10.1088/0004-637X/812/1/18}

\bibitem[{S.~L. {West} {et~al.}(2023){West}, {Lunnan}, {Omand}, {Kangas}, {Schulze}, {Strotjohann}, {Yang}, {Fransson}, {Sollerman}, {Perley}, {Yan}, {Chen}, {Chen}, {Taggart}, {Fremling}, {Bloom}, {Drake}, {Graham}, {Kasliwal}, {Laher}, {Medford}, {Neill}, {Riddle}, \& {Shupe}}]{2020qlb2023A&A...670A...7W}
{West}, S.~L., {Lunnan}, R., {Omand}, C.~M.~B., {et~al.} 2023, \bibinfo{title}{{SN 2020qlb: A hydrogen-poor superluminous supernova with well-characterized light curve undulations},} \aap, 670, A7, \dodoi{10.1051/0004-6361/202244086}

\bibitem[{S.~E. {Woosley}(2017){Woosley}}]{2017ApJ...836..244W}
{Woosley}, S.~E. 2017, \bibinfo{title}{{Pulsational Pair-instability Supernovae},} \apj, 836, 244, \dodoi{10.3847/1538-4357/836/2/244}

\bibitem[{L. {Yan} {et~al.}(2017){Yan}, {Lunnan}, {Perley}, {Gal-Yam}, {Yaron}, {Roy}, {Quimby}, {Sollerman}, {Fremling}, {Leloudas}, {Cenko}, {Vreeswijk}, {Graham}, {Howell}, {De Cia}, {Ofek}, {Nugent}, {Kulkarni}, {Hosseinzadeh}, {Masci}, {McCully}, {Rebbapragada}, \& {Wo{\'z}niak}}]{2017ApJ...848....6Y}
{Yan}, L., {Lunnan}, R., {Perley}, D.~A., {et~al.} 2017, \bibinfo{title}{{Hydrogen-poor Superluminous Supernovae with Late-time H{\ensuremath{\alpha}} Emission: Three Events From the Intermediate Palomar Transient Factory},} \apj, 848, 6, \dodoi{10.3847/1538-4357/aa8993}

\bibitem[{M. {Zechmeister} {\&} M. {K{\"u}rster}(2009){Zechmeister} \& {K{\"u}rster}}]{GLS2009A&A...496..577Z}
{Zechmeister}, M., \& {K{\"u}rster}, M. 2009, \bibinfo{title}{{The generalised Lomb-Scargle periodogram. A new formalism for the floating-mean and Keplerian periodograms},} \aap, 496, 577, \dodoi{10.1051/0004-6361:200811296}

\bibitem[{P. {Zhang} {et~al.}(2025){Zhang}, {Wang}, \& {Ji}}]{2025arXiv251209223Z}
{Zhang}, P., {Wang}, Z., \& {Ji}, S. 2025, \bibinfo{title}{{An extragalactic gamma-ray binary formed in supernova 2022jli},} arXiv e-prints, arXiv:2512.09223, \dodoi{10.48550/arXiv.2512.09223}

\bibitem[{J. {Zhu} {et~al.}(2023){Zhu}, {Jiang}, {Dong}, {Filippenko}, {Rudy}, {Pastorello}, {Ashall}, {Bose}, {Post}, {Bersier}, {Benetti}, {Brink}, {Chen}, {Dou}, {Elias-Rosa}, {Lundqvist}, {Mattila}, {Russell}, {Sitko}, {Somero}, {Stritzinger}, {Wang}, {Brown}, {Cappellaro}, {Fraser}, {Kankare}, {Moran}, {Prentice}, {Pursimo}, {Reynolds}, \& {Zheng}}]{2023ApJ...949...23Z}
{Zhu}, J., {Jiang}, N., {Dong}, S., {et~al.} 2023, \bibinfo{title}{{SN 2017egm: A Helium-rich Superluminous Supernova with Multiple Bumps in the Light Curves},} \apj, 949, 23, \dodoi{10.3847/1538-4357/acc2c3}

\end{thebibliography}

\end{document}